\documentclass[12pt]{article}
\pdfoutput=1

\usepackage{aas_macros,amsmath,amssymb,cite,esint,graphicx,mathtools,transparent}
\usepackage[margin=1in,letterpaper]{geometry}
\usepackage[colorlinks=true]{hyperref}
\usepackage[affil-it]{authblk}
\usepackage[normalem]{ulem}
\usepackage[compact]{titlesec}

\usepackage{epsfig}
\usepackage{amsfonts}
\usepackage{latexsym}
\usepackage{amsmath}
\usepackage{mathrsfs}
\usepackage{hyperref}
\usepackage{wasysym}
\usepackage{tikz}
\usepackage{caption}
\usepackage{subcaption}
\usepackage{float}
\numberwithin{equation}{section}

\DeclareFontFamily{OT1}{rsfs}{}
\DeclareFontShape{OT1}{rsfs}{m}{n}{
<-7> rsfs5 <7-10> rsfs7 <10-> rsfs10}{}
\DeclareMathAlphabet{\mycal}{OT1}{rsfs}{m}{n}

\newcommand{\bea}{\begin{align}}
\newcommand{\eea}{\end{align}}
\def\non{\nonumber}

\def\non{\nonumber}

\def\ads2{$\mathrm{AdS_2}$}
\def\nads2{$\mathrm{NAdS_2}$}
\def\sl2r{$\mathrm{SL(2,R)}$}
\def\tr{\tilde{r}}

\begin{document}

\unitlength = 1mm

\setcounter{tocdepth}{2}

\title{\textbf{\Huge Extreme lensing induces \linebreak {spectro-temporal correlations \linebreak in black-hole signals}}}

\date{}
\author[1,2]{Shahar~Hadar\thanks{shaharhadar@sci.haifa.ac.il}}
\author[2,3]{Sreehari~Harikesh\thanks{sharikes@campus.haifa.ac.il}}
\author[2,3]{Doron~Chelouche\thanks{doron@sci.haifa.ac.il}}

\affil[1]{\footnotesize Department of Mathematics and Physics, University of Haifa at Oranim, Kiryat Tivon 3600600, Israel
}
\affil[2]{\footnotesize Haifa Research Center for Theoretical Physics \& Astrophysics, University of Haifa, Haifa 3498838, Israel
}
\affil[3]{\footnotesize Department of Physics, Faculty of Natural Sciences, University of Haifa, Haifa 3498838, Israel
}

\maketitle

\vspace{-1.0cm}

\begin{abstract}
Rapid progress in electromagnetic black hole observation presents a theoretical challenge: how can the universal signatures of extreme gravitational lensing be distilled from stochastic astrophysical signals?
With this motivation, the two-point correlation function of specific intensity fluctuations across image positions, times, and frequencies is here considered. The contribution of strongly deflected light rays, those which make up the photon ring, is analytically computed for a Kerr black hole illuminated by a simple geometric-statistical emission model. We subsequently integrate over the image to yield a spectro-temporal correlation function which is relevant for unresolved sources. Finally, some observational aspects are discussed and a preliminary assessment of detectability with current and upcoming missions is provided.

\end{abstract}

\pagestyle{plain}
\setcounter{page}{1}
\newcounter{bean}
\baselineskip18pt

%%%%%%%%%%%%%%%%%%%%%%%%%%%%%%%%%%%%%%%%%%%%%%%%%%%%%%%%%%%%%%%%%%%%%

\setcounter{tocdepth}{2}

\section{Introduction}\label{sec:introduction}

The immediate surroundings of active black holes (BHs), at the horizon scale, provide an arena for some of the highest-energy astrophysical phenomena known to date. Plasma flowing into and around a BH at extreme temperatures, relativistic velocities, and strong magnetic fields, radiates through a variety of mechanisms and often becomes highly luminous. The emission generated in the BH's close vicinity is subsequently lensed by its strong gravitational field. Photons which eventually reach our telescopes, therefore, encode both the complicated astrophysical phenomena responsible for their creation, and the simple gravitational field governing their propagation. Nevertheless, the two effects are intertwined within the signal impinging on the telescope---which in many cases is also highly variable.

An outstanding challenge, therefore, is to find methods that disentangle the simple general relativistic (GR) effects from the complex astrophysical ones. In the context of interferometric imaging, it was proposed that the photon ring \cite{Darwin1959,Bardeen1973,Luminet1979,Falcke2000,Beckwith2005,Gralla2019,Johnson2020}, the image part resulting from extremely lensed\footnote{We use \emph{extreme lensing} to describe light ray deflection at large angles, in order to differentiate from `strong lensing' which customarily includes phenomena with small deflection angles such as Einstein rings.} light rays executing $n\geq1$ half-orbits around the BH, constitutes a universal component of the image and it therefore provides a probe of the BH's spacetime geometry. 
If optically thin, the unresolved ring's contribution is already fused into the remarkable images of M87* \cite{EHT2019a,EHT2019b,EHT2019c,EHT2019d,EHT2019e,EHT2019f} and SgrA* \cite{EHT2022a,EHT2022b,EHT2022c,EHT2022d,EHT2022e,EHT2022f}, released by the Event Horizon Telescope (EHT) collaboration in the past few years. Proposed multi-frequency space-very large baseline interferometry (VLBI) missions, e.g. EHE \cite{Kurczynski2022}, SALTUS \cite{Cardenas-Avendano2023}, and THEZA \cite{Gurvits2022}, are planned to dramatically decrease current EHT resolution to $\sim$ few $\mu \mathrm{as}$ in the future, and directly reveal the ring in time-averaged images.  
Efforts to detect the ring via superresolution techniques have been recently discussed in, e.g., \cite{Broderick2022,Lockhart2022,Tiede2022}.

Until future missions provide these transformative improvements in hardware capabilities, it could be worthwhile to simultaneously develop complementary techniques which probe extreme lensing with upcoming data. A possible alternative path towards probing the ring is to consider more general observables, in particular ones which exploit the sources' temporal variability. Ref.~\cite{Hadar2021} proposed an observable tailored to extract the GR information from stochastically varying emission near BHs: the two-point autocorrelation function of intensity fluctuations along the photon ring. Related proposed signatures include coherent light-curve autocorrelations \cite{Chesler2021}, BH glimmer \cite{Wong2020}, and imprints of localized hotspots, c.f. \cite{Broderick2006,Tiede2020,Cardoso2021,Emami2023}.

X-ray spectroscopy is a separate notable technique for BH observation. An especially prominent probe of GR is the broadening of emission lines, arising from monochromatic fluorescence of accretion flow ions, which is subsequently red/blueshifted by relativistic effects. Active galactic nuclei (AGN) spectroscopy has thus proven to be a fruitful means of estimating supermassive BH spin, see for example \cite{George1991,Brenneman2006,Reynolds2019,Gates2020a}, and the method is applicable to X-ray binaries (XRBs) as well, c.f. \cite{Bambi2017} and references therein.
Lagged correlation between different frequency bands has been explored in the context of X-ray reverberation mapping, see Ref.~\cite{Uttley2014} and references therein, where the echoes arise because of reflection off the accretion disk. For example, in \cite{Kara2016,Wilkins2016}, the lag between two broad bands (soft, $0.3-1$ keV; and hard, $1-4$ keV) was numerically studied. In a recent paper, Wilkins and collaborators report on observations consistent with light echoes of X-ray transients \cite{Wilkins2021} arising from reflection off the disk. 
Several past and present X-ray telescopes have measured AGN and XRB spectra with resolution $\sim \mathrm{few} \, 100$ eV (e.g., XMM, Suzaku, NuSTAR, AstroSat, NICER). 
The next-generation X-ray observatory NewAthena\footnote{the provisional name of the revised version of Athena, currently being reformulated at the ESA.} \cite{Athena_10FIRST_2013}
is expected to improve spectral resolution, collecting area, and other performance parameters by at least an order of magnitude. 

In this paper, we generalize the analysis of \cite{Hadar2021}, which considered flat-spectrum emission, to astrophysical sources with spectral dependence. We first introduce and compute the image spectro-temporal autocorrelation function (STAC), a two-point function which measures simultaneous correlations across different times, image positions and observation frequencies. This observable could in principle be relevant for multi-wavelength VLBI missions such as the next-generation EHT (ngEHT), which is set to add $86$~GHz and $345$~GHz to the current observation frequency of $230$~GHz. 
We introduce a geometric-statistical simplified model of a thin accretion disk with frequency-dependent emission in orbit around the BH, and analytically compute its image-STAC signature.
In order to model line emission, we then focus on sources which are monochromatic in the flow's rest frame. We analytically integrate over the image angles to obtain the unresolved-STAC function---an observable relevant for spatially unresolved, but spectrally and temporally resolved sources. We find a useful small-inclination expansion for both observables, which results in significant simplifications. Unresolved-STACs are significant since, if measurable in practice, they could probe extreme lensing in numerous unresolved sources.
Finally we provide a preliminary discussion of observational considerations, focusing on unresolved-STACs, and present an order of magnitude estimate of the signal-to-noise ratio (SNR) to be expected in such an observation.  

The central computation carried out in this paper involves a number of approximations and simplifying assumptions. Therefore, a more comprehensive study is required in order to assess its applicability, or to predict what the STAC functions will look like, in a more realistic setting. Natural follow-up questions are: what are the non-photon ring contributions, dominated by correlations of direct ($n=0$) and first subring ($n=1$) photons? Is our model good enough to characterize some realistic sources? Does current/upcoming data show STACs, and/or which improvements could allow their measurement? We thus suggest to view this work as a proposition for a direction of further study. The core idea, namely that the existence of a BH near a stochastic source induces correlations in observed signals \emph{is} robust and we view that as motivation to search for such signatures.

Following a brief review of the photon ring and shell in Sec.~\ref{sec:The Kerr photon shell & ring}, Sec.~\ref{sec:definition} surveys the observables considered in the paper. Subsequently, in Sec.~\ref{sec:calculation} we define our analytical toy model, which improves the one considered in \cite{Hadar2021} by including (a simplified model for) the flow velocity, and analytically compute the image-STAC function. We then integrate over image angles to obtain the unresolved-STAC function in Sec.~\ref{sec:unresolved STAC}, and discuss observational considerations in Sec.~\ref{sec:observation}. We use relativistic units $G=c=1$ unless stated otherwise.

\section{The Kerr photon shell \& ring}\label{sec:The Kerr photon shell & ring}

The Kerr geometry of mass $M$ and angular momentum $aM$ is described by the line element
\begin{gather}
\label{eq:Kerr line element}
    ds^2 = -\frac{\Delta}{\Sigma} \left( dt-a\sin^2\theta d\phi \right)^2 + \frac{\Sigma}{\Delta} dr^2 + \Sigma \, d\theta^2 + \frac{\sin^2\theta}{\Sigma} \left[ (r^2+a^2) d\phi -a dt \right]^2 \, ,  \\ \non \\
    \Delta = r^2-2Mr+a^2 \,, ~~~~~~~~~ \Sigma = r^2+a^2 \cos^2\theta  \, , \non 
\end{gather}
where $(t,r,\theta,\phi)$ are the standard Boyer-Lindquist coordinates. Geodesic motion in \eqref{eq:Kerr line element} is integrable due to the existence of four independent conserved quantities. In terms of the four-momentum $p^\mu$, these are the mass $\mu$, energy $-p_t$, azimuthal angular momentum $p_\phi$, and Carter constant $Q = p_\theta^2-\cos^2\theta[a^2(p_t^2-\mu^2)-p_\phi^2\csc^2\theta]$. For null geodesics $\mu=0$, the energy factors out and geodesics depend only on two impact parameters: the rescaled azimuthal angular momentum $\lambda=-p_\phi/p_t$, and the rescaled Carter constant $\eta=Q/p_t^2$.

The 6-dimensional phase space of (null) geodesics contains a special codimension-2 subspace of unstably bound, critical orbits at fixed Boyer-Lindquist radius $\tilde{r}_- \le \tilde{r} \le \tilde{r}_+$, with  
\begin{align}
    \tilde{r}_\pm = 2M\left[ 1+\cos\left( \frac{2}{3} \arccos (\pm a/M) \right) \right] \, ,
\end{align}
defined by the conditions
\begin{align} \label{eq:critical impact parameters}
    \lambda = \tilde{\lambda} = a+\frac{\tilde{r}}{a} \left[ \tilde{r} - \frac{2\tilde{\Delta}}{\tilde{r}-M} \right] \, , ~~~~~~~~~~ \eta = \tilde{\eta} = \frac{\tilde{r}^3 }{a^2} \left( \frac{4M \tilde{\Delta}}{(\tilde{r}-M)^2} -\tilde{r} \right)  \, ,
\end{align}
where we henceforth use tildes to indicate that quantities are evaluated precisely at criticality, e.g. $\tilde{\Delta}=\Delta(\tilde{r})$. The spacetime region spanned by all critical null geodesics is called the \emph{photon shell} \cite{Darwin1959,Bardeen1973,Luminet1979,Teo2003,Gralla2019,Johnson2020}. The behavior of a near-critical geodesic, which undergoes multiple half-orbits near the radius $\tilde{r}$, is characterized by three critical parameters: $\{\gamma(\tilde{r}),\delta(\tilde{r}),\tau(\tilde{r})\}$. The instability rate is characterized by the Lyapunov exponent $\gamma(\tilde{r})$ \cite{Johnson2020}: a near-critical geodesic approaching/receding from the critical radius $\tilde{r}$ which crosses the equatorial plane at $\{t_k,r_k,\phi_k\}$ after undergoing $k$ half-orbits, obeys $\delta r_{k+1}/\delta r_k \approx e^{\mp \gamma}$ where $\delta r_k = r_k - \tilde{r}$ is the coordinate deviation from the critical radius. The azimuthal and temporal durations of a half-orbit are given by $\phi_{k+1} - \phi_k \approx \delta + \delta \phi_k$, $t_{k+1}-t_k \approx \tau + \delta t_k$, where $\delta \phi_k,\delta t_k \to 0$ for $k \to \infty$ \cite{GrallaLupsasca2020a}. Exact analytical expressions for the critical parameters are given in App.~\ref{app:photon shell critical parameters}.

In order to describe the optical image of a BH, it is useful to introduce coordinates on the observer's screen. Here we are interested in distant observers $r_o \gg M$ at inclination $\theta_o$ relative to the BH spin axis, and we use the polar coordinates \cite{Bardeen1973,Johnson2020}
\begin{align} \label{eq:screen coordinates}
    \rho = \frac{1}{r_o} \sqrt{a^2 (\cos^2 \theta_o - u_+ u_-)+\lambda^2} \, , ~~~~~~~~~~~~ \cos \varphi = -\frac{\lambda}{r_o \rho \sin \theta_o} \, ,
\end{align}
where
\begin{align} \label{eq:u pm definitions}
    u_\pm = \Delta_\theta \pm \sqrt{\Delta_\theta^2 + \frac{\eta}{a^2}} \, , ~~~~~~~~~~~~ \Delta_\theta = \frac{1}{2} \left( 1-\frac{\eta + \lambda^2}{a^2} \right) \, .
\end{align}
The above polar screen coordinates are related to the standard Cartesian coordinates introduced by Bardeen via $(\alpha,\beta) = (\rho \cos \varphi,\rho \sin \varphi)$.
Evidently, a point on the screen defines a choice of impact parameters $(\lambda,\eta)$, and thus a particular light ray. The image of the (near-)critical orbits on the screen can be seen to define, via the critical impact parameters \eqref{eq:critical impact parameters}, a closed \emph{critical curve} $\tilde{\rho}(\varphi)$. This curve delineates light rays that begin at past null infinity from those which originate at the past horizon (for an eternal BH). Note also that \eqref{eq:critical impact parameters}, \eqref{eq:screen coordinates} define a relation $\tilde{r}(\varphi)$: remarkably, different angles around the critical curve are associated to different radii in the geometry.

\section{Simultaneous spectro-temporal autocorrelation}\label{sec:definition}

Ref.~\cite{Hadar2021} proposed to study 2-point autocorrelations of the specific intensity $I_{\nu}(t,\rho,\varphi)$ on a BH image at a particular fixed frequency, with EHT's observation frequency $\nu = 230 \mathrm{GHz}$ in mind. The source spectrum was taken to be flat, or frequency-independent \cite{Bower2015,Chael2019,EHT2019f}. In this paper we propose a generalization of this observable that could extract information from signals' frequency dependence. We first define the specific 2-point autocorrelation function on the image as
\begin{align}
\mathcal{C}^{3D}_{\nu \nu'}(T,\varphi,\varphi',\rho,\rho') = \langle \Delta I_\nu(t,\varphi,\rho) \Delta I_{\nu'}(t+T,\varphi',\rho') \rangle \, ,
\label{eq:C 3D def}
\end{align}
where $\Delta I \equiv I-\langle I \rangle$.  
Various integrals of \eqref{eq:C 3D def} produce interesting observables which may be applicable in different situations. Integrating with respect to image radii {\`a} la \cite{Hadar2021} yields
\begin{align}
\mathcal{C}^{2D}_{\nu \nu'}(T,\varphi,\varphi') = \int \rho d\rho \int \rho' d\rho' \mathcal{C}^{3D}_{\nu \nu'}(T,\varphi,\varphi',\rho,\rho')  \, .
\label{eq:C 2D def}
\end{align}
\eqref{eq:C 2D def} is a generalization of the 2-point photon ring autocorrelation function which includes frequency dependence. It could in principle be relevant for multi-wavelength VLBI, with ngEHT or future space-based missions. We compute it analytically for a simple model in Sec.~\ref{sec:calculation}. Integrating over the image angles as well gives
\begin{align}
\mathcal{C}_{\nu \nu'}(T) \equiv \mathcal{C}^{1D}_{\nu \nu'}(T) = \int d\varphi \int d\varphi' \mathcal{C}^{2D}_{\nu \nu'}(T,\varphi,\varphi') = \langle \Delta f_\nu(t) \Delta f_{\nu'}(t+T) \rangle \, ,
\label{eq:C 1D def}
\end{align}
the unresolved two-point STAC function of the specific flux fluctuations $\Delta f_\nu(t)$\footnote{\emph{Autocorrelation} is an appropriate term if $\Delta f_\nu(t)$ is viewed as a single function of two variables, $\nu$ and $t$, while \emph{cross-correlation} is suitable if the specific flux at each frequency band is (equivalently) viewed as a separate, single-variable function. Here we use the former nomenclature.}. The strong-lensing signatures of \eqref{eq:C 1D def} are computed in Sec.~\ref{sec:unresolved STAC}; its potential relevance for X-ray spectrometry is discussed in Sec.~\ref{sec:observation}. In the next section we compute $\mathcal{C}^{2D}_{\nu \nu'}(T)$ for a simple geometric-statistical emission model.

\section{Source model \& computation}\label{sec:calculation}

\begin{figure}[!ht]
	\centering
	\includegraphics[width=\textwidth]{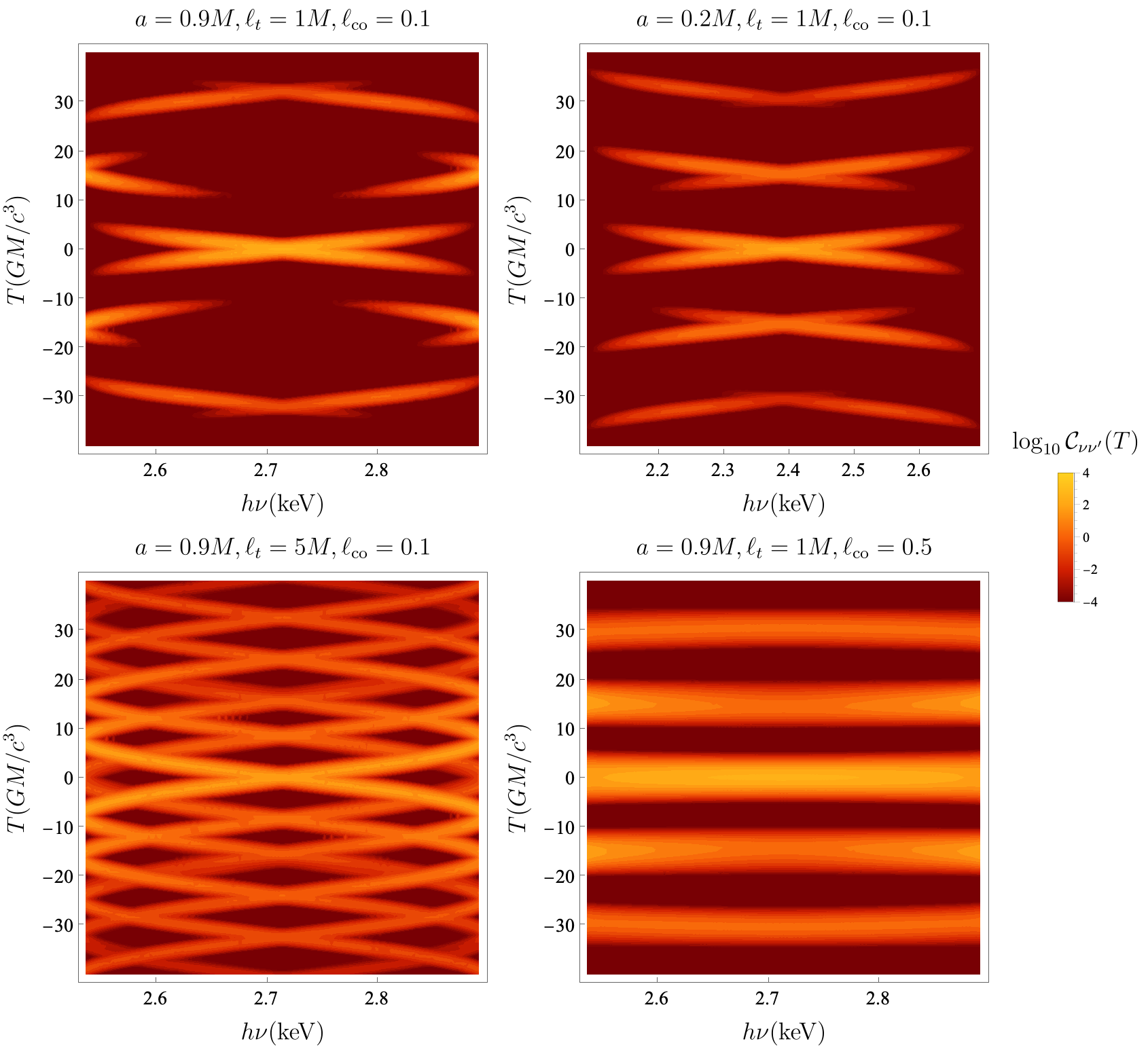}
	\caption{Quasi-universal contribution \eqref{eq:C1D result small inclination simple} to the spectro-temporal autocorrelation (STAC) function $C_{\nu \nu'}(T)$, in arbitrary units, for random line emission around a Kerr black hole of arbitrary mass $M$. We fix here $\nu'=\tilde{g}_0 \nu_{\mathrm{line}}$ and show the dependence on time delay $T$, in natural units, and on photon energy $h\nu$, where $h$ is Planck's constant. The black hole spin parameter is $a=0.2M$ in the upper right panel and $a=0.9M$ otherwise, and it is viewed at $11.5^\circ$ inclination with respect to the spin axis. The emitters are taken to move on equatorial circular orbits above the innermost stable circular orbit, and below it to plunge, while radiating monochromatically in their rest frame. The source's temporal correlation length is taken to be $\ell_t=5M$ in the bottom left panel and $\ell_t=1M$ otherwise, and its co-rotating azimuthal correlation length is taken to be $\ell_\mathrm{co}=0.5$ in the bottom right panel and $\ell_\mathrm{co}=0.1$ otherwise. In the emitters' rest frame the photon energy is taken to be $6.4$~keV, corresponding to the FeK$\alpha$ iron line. For such small inclination sources in the near-critical approximation \eqref{eq:near-critical brightness fluctuation}, nontrivial correlation structure appears only in the range given in \eqref{eq:nu pm}.}
	\label{fig:STACplotsb}
\end{figure}

\begin{figure}[!ht]
	\centering
	\includegraphics[width=\textwidth]{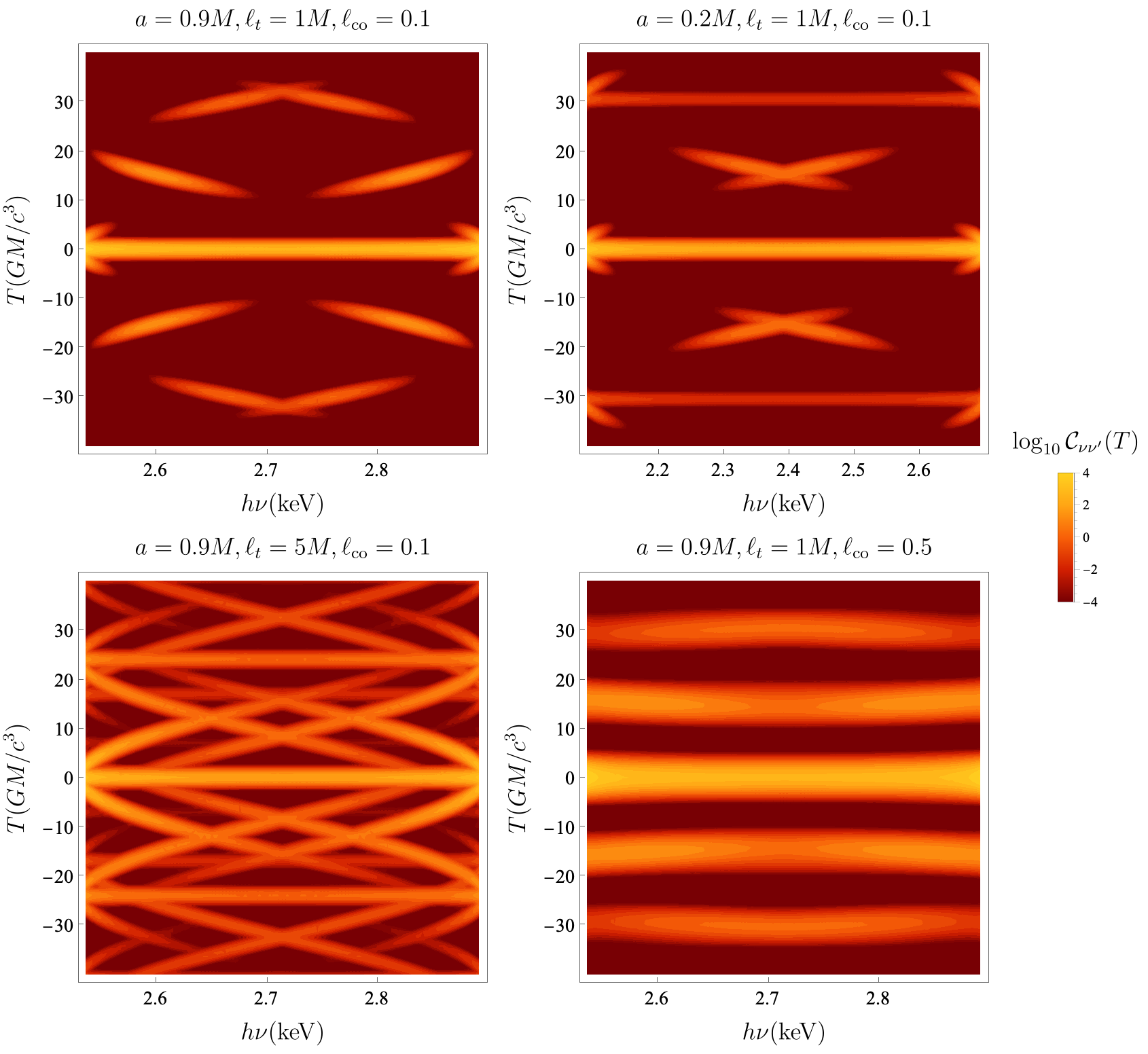}
	\caption{STAC function for the same setup as in Fig.~\ref{fig:STACplotsb}, this time fixing $\nu=\nu'$. The relative positions of correlation ridges are determined by the universal parameters $\tau_0$, $\delta_0$, the relative magnitudes by $\gamma_0$, their tilt by the ``quasi-universal'' accretion flow angular velocity $\Omega_s(\tilde{r}_0)$, and their widths by the statistical properties of the flow, modeled by $\ell_t$, $\ell_\mathrm{co}$; see Eq.~\eqref{eq:C1D result small inclination simple}.}
	\label{fig:STACplotsc}
\end{figure}

\begin{figure}[!ht]
	\centering
	\includegraphics[width=\textwidth]{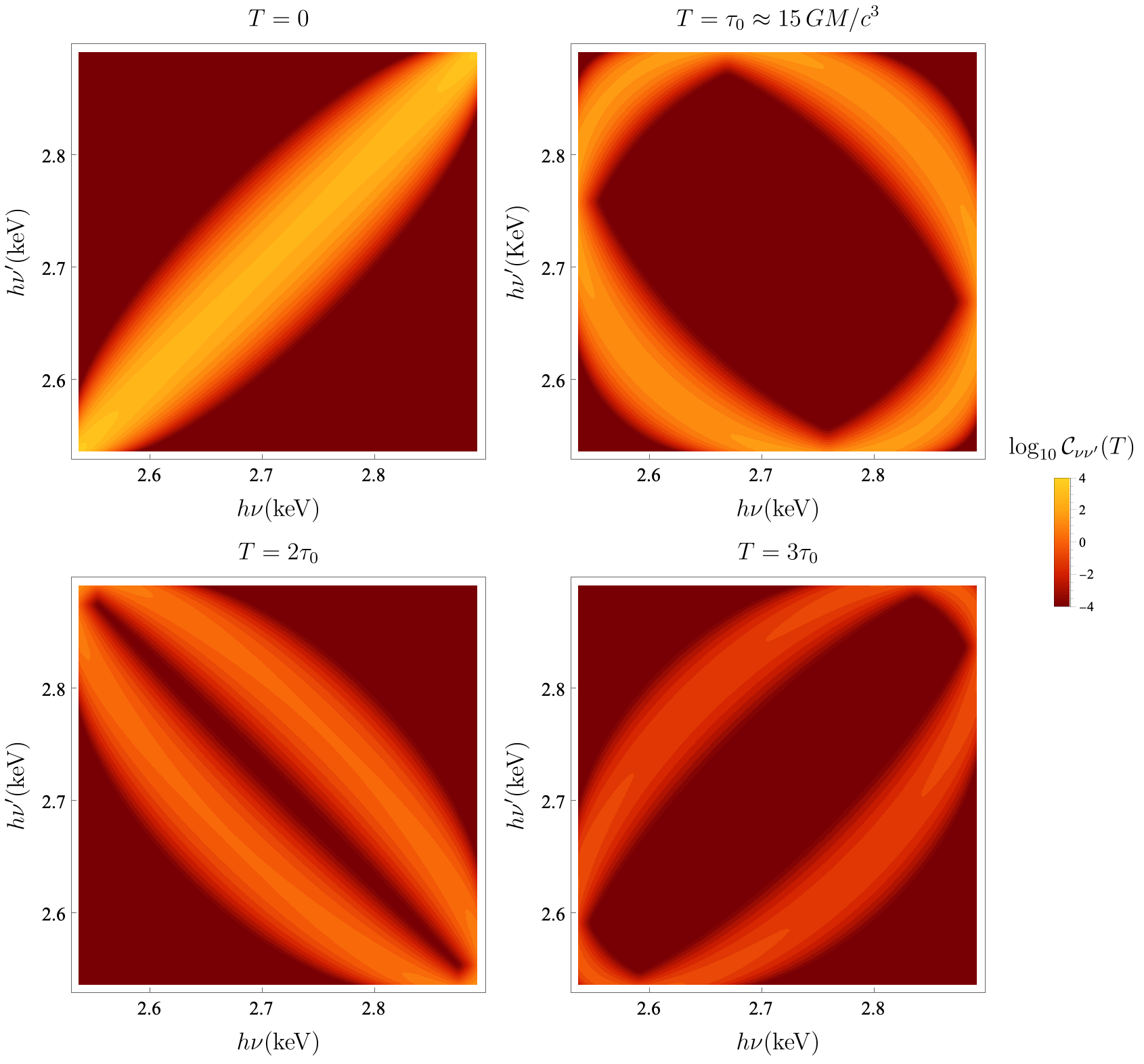}
	\caption{STAC function for the same setup as in Fig.~\ref{fig:STACplotsb}, with spin $a=0.9M$ and correlation lengths $\ell_t=1M$, $\ell_\mathrm{co}=0.1$ in all panels. Here, different fixed $T=0,\tau_0,2\tau_2,3\tau_0$ slices are displayed, showing the dependence on observed photon energies $\nu$, $\nu'$ in each slice.}
	\label{fig:STACplotsa}
\end{figure}

In this section we compute analytically the near-critical contribution to the image-STAC function \eqref{eq:C 2D def} for a simple stochastic emission model. Our model consists of a thin, equatorial disk composed of isotropic,
random emitters moving on circular orbits.
This amounts to considering a local emissivity profile of the form 
\begin{align}\label{eq:emissivity profile}
J_{\nu}(x^\alpha,\lambda,\eta) = \begin{cases} 
~~ S\left[\nu,g(r,\lambda,\eta)\right] \, \delta(\theta-\pi/2) j(t,r,\phi) &  ~~ r_{\mathrm{min}}<r<r_{\mathrm{max}} \\
~~ 0 & ~~~ \mathrm{else} 
\end{cases} \, ,
\end{align}
where $S$ encodes the frequency dependence, which varies both with the emission spectrum in the flow's rest frame and with the 
ratio of observed frequency $\nu_o$ to source rest-frame frequency $\nu_s$, known as the \emph{redshift factor}\footnote{$1<g<\sqrt{3}$ is possible in Kerr even for geodesic emitters, occurring whenever the Doppler blueshift overcomes the gravitational redshift. The upper bound $\sqrt{3}$ is saturated only in the maximally rotating case.}, given by
\begin{align}
    g \equiv \frac{\nu_o}{\nu_s} = \frac{u_{(o)}^\mu p^{(o)}_\mu}{u_{(s)}^\mu p^{(s)}_\mu} \, ,
\end{align}
where $u^\mu_{(s/o)}$ is the 4-velocity of the source/observer, and $p^{(s/o)}_\mu$ is the 4-momentum of the photon at the source/observer. $r_{\mathrm{max/min}}$ denote the outer/inner edges of the emission annulus satisfying $r_{\mathrm{max}}-r_\mathrm{min}=W$. By our assumptions, the redshift factor depends only on the emitter radius and the photon impact parameters. 
The emissivity's frequency dependence may be generally written as
\begin{align} \label{eq:source frequency dependence}
    S\left[\nu,g(r,\lambda,\eta)\right] = \int d\nu' \delta\left[ \nu- g(r,\lambda,\eta) \nu' \right] \sigma(\nu') \, ,
\end{align}
where $\sigma(\nu')$ models the emission spectrum in the flow's rest frame and $\delta$ is the Dirac delta function. 
For example, a particular choice which we will explicitly use in Sec.~\ref{sec:unresolved STAC}, is $\sigma(\nu) = \delta(\nu-\nu_\mathrm{line})$, or equivalently
\begin{align} \label{eq:monochromatic emitters}
    S\left[\nu,g(r,\lambda,\eta)\right] = \delta\left[ \nu- g(r,\lambda,\eta) \nu_\mathrm{line}\right] \, .
\end{align}
 This choice models line emission, i.e. a source which is monochromatic in its rest frame, with emission frequency $\nu_\mathrm{line}$\footnote{For concreteness, an example we use in Figs.~\ref{fig:STACplotsb}, \ref{fig:STACplotsc}, \ref{fig:STACplotsa}, is the FeK$\alpha$ line at $\nu_\mathrm{line} \approx 6.4 \mathrm{keV}$.}. Note that in the Boyer-Lindquist static frame the emission \eqref{eq:emissivity profile} is neither monochromatic nor isotropic even when Eq.~\eqref{eq:monochromatic emitters} holds.

The redshift $g$ depends on the source's position and velocity\footnote{and generally on the observer's as well, but we specify here to a stationary asymptotic observer. }. As a concrete example, we will sometimes use an explicit formula which holds for emitters moving along geodesic equatorial circular orbits (GECO) at angular velocity $\Omega_s = \partial_t \phi_s = \frac{\pm \sqrt{M}}{r_s^{3/2}\pm a\sqrt{M}}$, where $\pm$ is used for co-/counter-rotating orbits, respectively. The redshift is then simply expressed as a function of emission radius $r_s$ and azimuthal impact parameter $\lambda$ \cite{Cunningham1975}: 
\begin{align}
g_{\mathrm{GECO}} = \frac{\sqrt{r^3_s-3M r_s^2 \pm 2a\sqrt{M}r_s^{3/2}}}{r_s^{3/2} \pm \sqrt{M}(a-\lambda)} \, .
\label{eq:g for equatorial circular geodesic emitters}
\end{align}
Below the innermost stable circular orbit (ISCO), initially circular geodesics plunge into the horizon with approximately ISCO angular momentum. An exact expression \cite{Cunningham1975} for the redshift factor $g_\mathrm{plunge}(r_s,\lambda)$ of such emitters is reviewed in App.~\ref{app:Redshift}. This expression can be used to model the innermost, plunging portion of such a circularized thin geodesic disk by assuming $g=g_{\mathrm{orbit}}$ for $r \geq r_{\mathrm{isco}}$ and $g=g_{\mathrm{plunge}}$ for $r < r_{\mathrm{isco}}$. Different choices for the average brightness (or the standard deviation of its fluctuations) at $r < r_{\mathrm{isco}}$ can model flows with different accretion rates. Here we choose the same standard deviation for brightness fluctuations at all relevant disk radii, for simplicity; see equation \eqref{eq:J 2PF} below.
Assuming that correlations of the emissivity fluctuations $\Delta j = j - \langle j \rangle$ are local, and stationarity and axisymmetry on average, we model the 2-point correlation function of $\Delta j$ as
\begin{align}\label{eq:J 2PF}
\langle \Delta j(t,r,\phi) \Delta j(t',r',\phi') \rangle = \mathcal{J} G_{\ell_t}(t-t') G_{\ell_r} (r-r') G^\circ_{\ell_\mathrm{co}}\left[\phi-\phi'-\Omega_s\left(\frac{r+r'}{2}\right) \, (t-t') \right] \, ,
\end{align}
for $r_{\mathrm{min}}<r,r'<r_{\mathrm{max}}$, where
\begin{align}
G_{\ell}(z) = \frac{1}{\sqrt{2 \pi} \, \ell} e^{-\frac{z^2}{2\ell^2}} \, ,~~~ G^{\circ}_{\ell}(\phi) = \frac{1}{2\pi I_0(1/\ell^2)e^{-1/\ell^2}}e^{-\frac{2}{\ell^2}\sin^2\left( \phi/ 2\right)} \, ,
\end{align}
are the Gaussian and its analogue for a periodic variable, the von Mises distributions, respectively; $I_0(z)$ denotes the modified Bessel function. 
Eq.~\eqref{eq:J 2PF} improves the model of \cite{Hadar2021} by requiring emissivity correlations to be most localized in the flow's rest frame.
Note that in \eqref{eq:emissivity profile},\eqref{eq:J 2PF} we have assumed for simplicity that the frequency dependence factors out of the emissivity fluctuations' two-point function, and
that the velocity field is homogeneous, i.e. that velocity fluctuations in the flow can be neglected when determining the red/blueshift associated with a specific location in the disk. Whenever the latter assumption fails to hold, one expects an additional stochastic broadening of the spectrum already in the flow's average rest frame. It would be interesting to generalize by relaxing the above two assumptions.

Due to extreme lensing, a light ray arriving at the observer screen may intersect the emission region multiple times; for a light ray with half-orbit number $n$, the specific intensity fluctuation on the screen at observation frequency $\nu$ is given by \cite{Hadar2021}
\begin{align}
\Delta I_{\nu}(t,\rho,\varphi) =\sum_{k=0}^{n} \Delta I^{(k)}_{\nu} = \sum_{k=0}^{n} g_{(k)}^3 S\left(\nu,g_{(k)}\right) \, \frac{r_s^{(k)} }{\cos\Theta_{(k)}} \, \Delta j(t_s^{(k)},r_s^{(k)},\phi_s^{(k)}) \, ,
\end{align}
where $\Delta I^{(k)}_{\nu_o}$ is the contribution from the $k_{\mathrm{th}}$ intersection of the source, $g_{(k)}=g_{(k)}(r_{(k)},\rho,\varphi)=g_{(k)}(r_{(k)},\lambda,\eta)$ is the redshift factor, and $\cos\Theta_{(k)}$ is the emission angle cosine relative to the local zenith in the source rest frame.
For example, for corotating GECOs, $\cos\Theta=\pm g \sqrt{\eta}/r$ where the $\pm$ sign corresponds to emission upwards/downwards from the disk.
$x_s^{(k)}(\rho,\varphi)$ for $x \in \{t,r,\phi\}$ describe the location of the $k_{\mathrm{th}}$ intersection.

We now focus on near-critical light rays. These impinge on the screen at $\rho=\tilde{\rho}(\varphi)+\delta \rho$ with $\delta \rho/\tilde{\rho}\ll1$, and their half-orbit number scales as $n \sim -\ln \left( \delta \rho/\tilde{\rho}\right)$. For such light rays we may apply the approximations reviewed in section \ref{sec:The Kerr photon shell & ring}. 
The redshift and emission angle cosine in the emitter's rest frame converge exponentially with $k$ to their critical value as the geodesic moves into the ``near-photon shell' regime \cite{Hadar2022}. Therefore, for large enough $k<n$ we may approximate $g_{(k)}\approx \tilde{g}=g(\tilde{r},\lambda(\tilde{r}),\eta(\tilde{r}))$, $\cos\Theta_{(k)} \approx \cos\tilde{\Theta}$, and these become $k$-independent functions of the screen angle $\varphi$. In practice, the lowest $k$ from which this approximation becomes good depends on details of the system---BH spin and observer inclination, the accretion flow's velocity field and statistics---as well as the desired level of approximation. We will denote the lowest value above which the photon shell approximation applies by $k_\mathrm{ps}$. In practice, it seems like $k_\mathrm{ps}=\{1,2\}$ could be reasonable choices in different situations, but more study of the low $k$ contributions is required in order to justify a particular choice and quantify the non-universal correlations. 

In this paper we focus on the $k\geq k_\mathrm{ps}$ ``quasi-universal''\footnote{We use the term \emph{quasi-universal} since the redshift relation $g(r,\lambda,\eta)$ generally depends on accretion flow parameters, e.g. magnetic field structure. The simple case of circular GECOs \eqref{eq:g for equatorial circular geodesic emitters} is a consequence of GR alone and models a geometrically thin disk.} contribution to STACs. 
Under the above-described assumptions, we express the brightness fluctuations near the critical curve as
\begin{align}\label{eq:near-critical brightness fluctuation}
\Delta &I_{\nu}(t,\tilde{\rho}+\delta \rho,\varphi) \approx \sum_0^{k_\mathrm{ps}-1} \Delta I^{(k)}_{\nu} + \frac{\tr \tilde{g}^3 S\left(\nu,\tilde{g}\right) }{\cos \tilde{\Theta}} \\ &\times \sum_{k=k_\mathrm{ps}}^{n} \, \Delta j\left[t-\Delta t_{\pm_\beta} \left( \tr \right)-k \tau\left( \tr \right), \tr+\delta r_k,\phi_o-\Delta \phi_{\pm_\beta} \left( \tr \right)-k \delta\left( \tr \right)\right] \, , \non
\end{align}
where $\tr=\tr(\varphi)$---a relation determined by Eqs.~\eqref{eq:critical impact parameters}, \eqref{eq:screen coordinates}---$\phi_o$ is the observer azimuthal angle, and $\pm_{\beta}=\mathrm{sign}(\beta)=\mathrm{sign}(\sin \varphi)$.
Note that all quantities with tilde are a function of $\varphi$ alone via their dependence on photon shell radius, e.g. $\tilde{g}=g(\tr(\varphi))$ (and on $\pm_\beta$ when applicable).
$\Delta t$ ($\Delta \phi$) denote the $\sim k^0$ piece of the total time (azimuth) traversed between source and observer, which is formally subleading in a large-$k$ expansion.

Next, we plug \eqref{eq:near-critical brightness fluctuation} into the definition \eqref{eq:C 2D def}, use \eqref{eq:J 2PF},
exchange the order of summation and integration, and perform the radial screen integrals\footnote{For an elaborate account of the computation of such integrals, see \cite{Hadar2021}.}. The quasi-universal part of the result of this computation, which is composed of $k,k' \geq k_\mathrm{ps}$ contributions, is
\begin{align} \label{eq:C2D result}
&\mathcal{C}^{2D}_{\nu \nu'}(\varphi,\varphi',T) = N(\varphi,\varphi') S\left(\nu,\tilde{g}(\varphi)\right) S\left(\nu',\tilde{g}(\varphi')\right) \non \\
& \times \sum_{k=k_\mathrm{ps}}^\infty \sum_{k'=k_\mathrm{ps}}^\infty e^{-k\gamma\left(\tr(\varphi) \right)} e^{-k'\gamma\left(\tr(\varphi')\right)} G_{\ell_t}\left[ T+\Delta t_{\pm_{\beta}} \left(\tr(\varphi)\right)-\Delta t_{\pm_{\beta'}} \left(\tr(\varphi')\right) +k \tau \left( \tr(\varphi)\right)-k' \tau \left( \tr(\varphi')\right) \right] \non \\  &\times G^\circ_{\ell_\mathrm{co}}\left[ \Delta \phi_{\pm_{\beta}} \left(\tr(\varphi)\right)-\Delta \phi_{\pm_{\beta'}} \left( \tr(\varphi') \right) +k \delta \left( \tr(\varphi)\right) - k' \delta \left(\tr(\varphi')\right) \right. \non \\
& ~~~~~~~~~~~~~ \left. -\Omega_s\left(\frac{\tr(\varphi)+\tr(\varphi')}{2}\right)\left( T+\Delta t_{\pm_{\beta}} \left(\tr(\varphi)\right)-\Delta t_{\pm_{\beta'}} \left(\tr(\varphi')\right) +k \tau \left( \tr(\varphi)\right)-k' \tau \left( \tr(\varphi')\right) \right) \right] \, ,
\end{align}
where
\begin{align}
 N(\varphi,\varphi') =  \frac{\mathcal{J} \tilde{r} \tilde{r}' \tilde{\rho} \tilde{\rho}' \tilde{\iota}_{\pm_\beta} \tilde{\iota}'_{\pm_\beta} \tilde{g}^3 \tilde{g}'^3}{\cos \tilde{\Theta} \cos \tilde{\Theta}'} \, W \Lambda\left( \frac{\tilde{r}-\tilde{r}'}{W} \right) \, ,
\end{align}
with $\tr=\tr(\varphi)$ and $\tr'=\tr(\varphi')$ implied, and
\begin{align}
    &\Lambda_{\ell/W}(z) := \frac{1+z}{2} \mathrm{erf} \left( \frac{1+z}{\sqrt{2}\ell/W} \right) - \frac{z}{2} \mathrm{erf} \left( \frac{z}{\sqrt{2}\ell/W} \right) + \left( \ell/W \right)^2 \left( e^{-\frac{z+1/2}{\left( \ell/W \right)^2}} -1 \right) G_{\ell/W}(z) \non \\
    &+(z \to -z) 
    \, ,
\end{align}
\begin{align}
\frac{1}{\iota_{\pm_\beta}(\tr)}:= \frac{1+\sqrt{\tilde{\chi}}}{32 \tilde{r}^3 \tilde{\chi}^2} \sqrt{\tilde{\beta}^2+\tilde{\psi}^2} \Delta(\tilde{r}) e^{\mp_{\beta}2\tilde{r}\sqrt{\tilde{\chi}}f_o} \sin \tilde{\Psi} \, ,
\end{align}
and
\begin{align}
    & \tilde{\psi} = \tilde{\alpha} - \frac{\tilde{r}+M}{\tilde{r}-M} a \sin \theta_o \, , ~~~~~~~ 
    \tilde{\chi} = 1-\frac{M \Delta(\tilde{r})}{\tilde{r} (\tilde{r}-M)^2} \, , \non \\ \non \\
    & \cot \tilde{\Psi}(\varphi) = \partial_\varphi \ln \tilde{\rho}(\varphi) \, ,  ~~~~~~~ 
    f_o = \frac{1}{a \sqrt{-\tilde{u}_-}} F\left( \arcsin \frac{\cos \theta_o}{\sqrt{\tilde{u}_+}} \middle| \frac{\tilde{u}_+}{\tilde{u}_-}  \right) \, ,
\end{align}
where $F$ is the incomplete elliptic integral of the first kind.

\subsection{Expansion in small inclination} \label{sec: small inclination approximation}
For small observer inclination $\sin \theta_o \ll 1$, \eqref{eq:C2D result} significantly simplifies. Since only $\lambda=0$ geodesics can precisely reach the pole $\theta=0$, inverting \eqref{eq:critical impact parameters} yields a single photon shell radius accessible for a polar observer, given by
\begin{align}
    \tilde{r}_0 = M+2\sqrt{M^2-\frac{a^2}{3}} \cos \left[ \frac{1}{3} \arccos \left( \frac{1-\frac{a^2}{M^2}}{\left( 1-\frac{a^2}{3M^2} \right)^{3/2}} \right) \right] \, .
\end{align}  
For $\sin\theta_o \ll 1$, the photon shell radii
\begin{align}
    \tilde{r}(\varphi) \approx \tilde{r}_0 + \Xi \, a \sin \theta_o \cos \varphi \, ,
\end{align}
become accessible, where 
\begin{align}
    \Xi = \left( \frac{\Delta(\tilde{r}_0)}{\tilde{r}_0-M} -M \right) \, \frac{4 \tilde{r}^2_0 \sqrt{\Delta(\tilde{r}_0)}}{3M^2 (\tilde{r}_0^2+a^2)+a^2\left(\Delta(\tilde{r}_0)-6M \tilde{r}_0\right)} \, .
\end{align}
The small inclination approximation for $Y \in \left\{ \tilde{g}, \gamma, \delta, \tau, \cos{\tilde{\Theta}, \tilde{\rho}, \Omega_s}  \right\}$ is given by
\begin{align}\label{eq:small inclination expansion of even quantities}
    Y=Y_0+Y_1 a \sin \theta_o \cos \varphi \, , ~~~~~~~ Y_1 = \Xi \, \partial_{\tilde{r}} \left. Y \right|_{\tilde{r} = \tilde{r}_0} \, ,
\end{align}
while for $Z \in \left\{\iota, \Delta t, \Delta \phi \right\}$, both even and odd parts in $\varphi$ exist because of the dependence on $\pm_\beta$, and the expansion is given by
\begin{align}
    Z=Z_0+ \sin \theta_o \left( a Z^{e}_1 \cos \varphi + Z^{o}_1 \sin \varphi \right) \, \, ,
\end{align}
where $Z^{e}_1$ is defined analogously to $Y_1$ in \eqref{eq:small inclination expansion of even quantities}, and $Z^{o}_1$ quantifies the odd-parity $\varphi$-dependence and is obtained by deriving $Z$ with respect to $\varphi$ while holding $\tr$ fixed.
Note that for a generic source velocity profile the leading small-inclination correction to the redshift is $a$-independent, as Doppler blue/redshift occurs due to orbital motion. This indeed is the case for GECOs; Eq.~\eqref{eq:g for equatorial circular geodesic emitters} shows that $\tilde{g}_1 \sim 1/a$ for low spin.

Following \cite{Hadar2021}, where details may be found, we expand \eqref{eq:C2D result} in small inclination up to $\mathcal{O}(\sin\theta_o)$ (inclusive). This allows to perform the sum over $k+k'$ and thus only a sum over $m=k-k'$ remains. The final result of this computation is
\begin{align} \label{eq:C2D result small inclination}
    &\mathcal{C}_{\nu\nu'}(\varphi,\varphi',T) \approx S\left(\nu,\tilde{g}(\varphi)\right) S\left(\nu',\tilde{g}(\varphi')\right) \frac{c\left( \varphi,\varphi' \right)}{1-e^{-2\gamma_0}} e^{-2k_\mathrm{ps} \gamma_0} \non \, \\
    &\times \sum_{-m_\mathrm{max}}^{m_\mathrm{max}} e^{- |m| \gamma_0} \left( 1-a\sin\theta_o B_m \right) G_{\ell_t}\left(T+m\tau_0\right) G^\circ_{\ell_\mathrm{co}}\left(\varphi'-\varphi+m\delta_0 - \Omega_s(\tr_0) \left( T+m \tau_0 \right) \right) \, ,
\end{align}
where $m_\mathrm{max}$ is the largest $m$ s.t. $|m| \tau_1 a \sin \theta_o \ll \ell_t $, $|m| \delta_1 a \sin \theta_o \ll \ell_\phi $, $|m| \gamma_1 a \sin \theta_o \ll 1 $,
\begin{align}
    &c\left( \varphi,\varphi' \right) = 
    \mathcal{J} W \Lambda(0) \left( \frac{\tilde{g}_0^3 \tilde{r}_0\tilde{\rho}_0 \iota_0}{\cos\tilde{\Theta}_0} \right)^2 \\  & \times \left[ 1 +a\sin\theta_o \left( \cos\varphi + \cos\varphi' \right) \left( 3\frac{\tilde{g}_1}{\tilde{g}_0} +\frac{\Xi}{\tilde{r}_0}+\frac{\tilde{\rho}_1}{\tilde{\rho}_0}+\frac{\iota^e_1}{\iota_0}-\frac{\cos\tilde{\Theta}_1}{\cos\tilde{\Theta}_0} \right) +\sin\theta_o \frac{\iota^o_1}{\iota_0} \left( \sin\varphi + \sin\varphi' \right)  \right] \, , \non
\end{align}
and
\begin{align}
    &2 B_m\left( T, \varphi,\varphi' \right) = \left( \cos\varphi - \cos\varphi' \right) \left[ -m\gamma_1 +  \frac{T+m \tau_0}{\ell_t^2} \left( 2 \Delta t^e_1 + \tau_1 \Sigma_m  \right)\right. \non \\
    & \left. + \frac{\sin\left( \varphi'-\varphi+m\delta_0 - \Omega_s^0 \left( T+m \tau_0 \right) \right)}{ \ell_\mathrm{co}^2} \left( 2 \left(\Delta \phi^e_1 - \Omega_s^0 \Delta t^e_1 \right) + \left(\delta_1 - \Omega_s^0 \tau_1 \right) \Sigma_m \right) \right] \non \\
    &  + \frac{2}{a}  \left( \sin\varphi - \sin\varphi' \right)  \left[ \frac{T+m \tau_0}{\ell_t^2} \Delta t^o_1  +\frac{\sin\left( \varphi'-\varphi+m\delta_0 - \Omega_s^0 \left( T+m \tau_0 \right) \right)}{ \ell_\mathrm{co}^2} \left(\Delta \phi^o_1 - \Omega_s^0 \Delta t^o_1 \right)\right]    \non \\
    & + \left( \cos\varphi + \cos\varphi' \right) \left[ \gamma_1 \Sigma_m +  \frac{T+m \tau_0}{\ell_t^2} m \tau_1 \right. \non \\ 
    & \left. ~~~~~~~~~~~~~~~~~~~~~~~~~ + \frac{\sin\left(\varphi'-\varphi+m\delta_0 - \Omega_s^0 \left( T+m \tau_0 \right) \right)}{\ell_\mathrm{co}^2} \left(  m \left(\delta_1 - \Omega_s^0 \tau_1 \right) -\Omega_s^1 (T+m\tau_0) \right) \right] \, , \non \\
\end{align}
where $\Sigma_m= \frac{2}{e^{2\gamma_0}-1}+|m|+2k_\mathrm{ps}$.

It is straightforward to check that integrating \eqref{eq:C2D result small inclination} with respect to $\nu,\nu'$ and taking $\Omega_s \to 0$ agrees with \cite{Hadar2021}, where the flat spectrum case was considered. The location of the correlation peaks in the 2D plane spanned by $\left( \varphi-\varphi',T \right)$ is controlled by $\tau_0,\delta_0$, universally depending on BH parameters only. The most prominent effect of a nonzero $\Omega_s$ in \eqref{eq:C2D result small inclination}, whenever $\ell_t,\ell_{\mathrm{co}}$ are small enough compared to $\tau_0$ and $\delta_0$,
is to tilt the orientation of the peaks in the $\left( \varphi-\varphi',T \right)$ plane. In the next section we will specify to line emission and integrate our results over image angles, thereby obtaining observables relevant for unresolved sources.

\section{Spatially unresolved sources with line emission} \label{sec:unresolved STAC}

It is interesting to integrate $\mathcal{C}^{2D}_{\nu \nu'}(T,\varphi,\varphi')$ \eqref{eq:C 2D def} over the angles $\varphi$, $\varphi'$. The main advantage of the resulting STAC function $\mathcal{C}_{\nu \nu'}(T)$ \eqref{eq:C 1D def}, is that it is potentially useful for spatially unresolved sources, as long as they are resolved well enough spectrally. 
With an application for line emission in mind, we specify henceforth to sources which are monochromatic in their rest frame, see Eq.~\eqref{eq:monochromatic emitters}.
Integrating the quasi-universal contribution \eqref{eq:C2D result} computed for the simple model considered in Sec.~\ref{sec:calculation} over angles removes the Dirac delta functions there. Since $\tr(-\varphi) = \tr(\varphi)$, for $0<|\varphi|<\pi$ the argument of each delta function has two zeros, at $\nu=\tilde{g}(\pm\varphi)\nu_\mathrm{line}$. In the range $\nu_- < \nu,\nu' < \nu_+ $ where $\nu_- = \tilde{g}(\varphi=0)\nu_\mathrm{line}$ and $\nu_+ = \tilde{g}(\varphi=\pi)\nu_\mathrm{line}$, the result is
\begin{align}
&\mathcal{C}_{\nu \nu'}(T) = \int d\varphi \int d\varphi' \, \mathcal{C}^{2D}_{\nu \nu'}(\varphi,\varphi',T) =   \sum_{\pm_\beta , \pm_{\beta'}} \frac{N\left( \varphi_{\pm_{\beta}}(\nu),\varphi_{\pm_{\beta'}}(\nu') \right)}{\nu_\mathrm{line}^2 \left|\frac{d\tilde{g}}{d\varphi}(\nu) \frac{d\tilde{g}}{d\varphi}(\nu')\right|}   \\
& \times \sum_{k=k_\mathrm{ps}}^\infty \sum_{k'=k_\mathrm{ps}}^\infty  e^{-k\gamma\left(\tr(\nu) \right)} e^{-k'\gamma\left(\tr(\nu')\right)} G_{\ell_t}\left[ T+\Delta t_{\pm_\beta} \left(\tr(\nu)\right)-\Delta t_{\pm_{\beta'}} \left(\tr(\nu')\right) +k \tau \left( \tr(\nu)\right)-k' \tau \left( \tr(\nu')\right) \right] \non \\  &\times G^\circ_{\ell_\mathrm{co}}\left[ \Delta \phi_{\pm_\beta} \left(\tr(\nu)\right)-\Delta \phi_{\pm_{\beta'}} \left( \tr(\nu') \right) +k \delta \left( \tr(\nu)\right) - k' \delta \left(\tr(\nu')\right) \right. \non \\
& ~~~~~~~~~~~~~ \left. -\Omega_s\left(\frac{\tr(\nu)+\tr(\nu')}{2}\right)\left( T+\Delta t_{\pm_\beta} \left(\tr(\nu)\right)-\Delta t_{\pm_{\beta'}} \left(\tr(\nu')\right) +k \tau \left( \tr(\nu)\right)-k' \tau \left( \tr(\nu')\right) \right) \right] \notag \, .
\label{eq:C1D result frequency}
\end{align}
The dependence on $\nu$, $\nu'$ and $\pm_\beta$, $\pm_{\beta'}$ arises from the relation imposed by the delta functions \eqref{eq:monochromatic emitters}, $\varphi_{\pm_{\beta}}(\nu) = \pm_\beta \, \tilde{g}^{-1}(\nu/\nu_\mathrm{line})$, where $\tilde{g}^{-1}$ denotes the inverse function of $\tilde{g}(\varphi)$.
Recall that e.g. for GECOs, taking $r_s=\tr$ and $\lambda=\tilde{\lambda}$ in Eq.~\eqref{eq:g for equatorial circular geodesic emitters} defines $\tilde{g}(\varphi)$. 
Since $\tilde{g}(\varphi)$ is not injective, there are two possible ways to invert this function, corresponding to $\pm_\beta$. Note that the first sum runs over $\pm_{\beta}=\pm1$, $\pm_{\beta'}=\pm1$ and thus yields four different terms.

It is interesting to inspect the small inclination limit of Eq.~\ref{eq:C1D result frequency}, or equivalently to integrate \eqref{eq:C2D result small inclination} over $\varphi,\varphi'$. In this case, the angle-frequency relation \eqref{eq:small inclination expansion of even quantities} may be simply inverted explicitly,
\begin{align}
    \varphi_{\pm_{\beta}}(\nu) = \pm_{\beta} \arccos \frac{\nu/\nu_{\mathrm{line}}-\tilde{g}_0}{a \sin \theta_o \tilde{g}_1} \, .
\end{align}
The result up to $\mathcal{O}(\sin\theta_o)$ (inclusive) is given, in the range $\nu_-<\nu,\nu'<\nu_+$ where \begin{align}\label{eq:nu pm}
    \nu_\pm = \nu_{\mathrm{line}} (\tilde{g}_0 \pm a \sin \theta_o \tilde{g}_1) \, ,
\end{align} 
by
\begin{align} \label{eq:C1D result small inclination}
    &\mathcal{C}_{\nu\nu'}(T) \approx 
    \sum_{\pm_\beta , \pm_{\beta'}} \frac{c\left[ \varphi_{\pm_{\beta}}(\nu),\varphi_{\pm_{\beta'}}(\nu') \right]}{\nu_\mathrm{line}^2 \sqrt{(a \sin \theta_o \tilde{g}_1)^2 - (\nu/\nu_{\mathrm{line}}-\tilde{g}_0)^2} \sqrt{(a \sin \theta_o \tilde{g}_1)^2 - (\nu'/\nu_{\mathrm{line}}-\tilde{g}_0)^2}}
    \frac{e^{- 2k_\mathrm{ps} \gamma_0}}{1-e^{-2\gamma_0}} \non \, \\
    &\times \sum_{-m_\mathrm{max}}^{m_\mathrm{max}} e^{- |m| \gamma_0} \left\{ 1-a\sin\theta_o B_m \left[T, \varphi_{\pm_{\beta}}(\nu),\varphi_{\pm_{\beta'}}(\nu') \right] \right\} \, G_{\ell_t}\left(T+m\tau_0\right) \\
    & \qquad \times G^\circ_{\ell_\mathrm{co}}\left[\pm_{\beta'} \arccos \frac{\nu'/\nu_{\mathrm{line}}-\tilde{g}_0}{a \sin \theta_o \tilde{g}_1} \mp_{\beta} \arccos \frac{\nu/\nu_{\mathrm{line}}-\tilde{g}_0}{a \sin \theta_o \tilde{g}_1} +m\delta_0 - \Omega_s(\tr_0) \left( T+m \tau_0 \right) \right]  \notag \, .
\end{align}
In order to gain intuition for the above result we can choose to neglect the $\propto \sin \theta_o$ corrections, for clarity, and obtain a simpler expression, accurate to $\mathcal{O}(\sin \theta^0_o)$:
\begin{align} \label{eq:C1D result small inclination simple}
    &\mathcal{C}_{\nu\nu'}(T) \approx 
    \sum_{\pm_\beta , \pm_{\beta'}} \frac{\mathcal{J} W \Lambda(0) \left( \frac{\tilde{g}_0^3 \tilde{r}_0\tilde{\rho}_0 \iota_0}{\cos\tilde{\Theta}_0} \right)^2}{\nu_\mathrm{line}^2 \sqrt{(a \sin \theta_o \tilde{g}_1)^2 - (\nu/\nu_{\mathrm{line}}-\tilde{g}_0)^2} \sqrt{(a \sin \theta_o \tilde{g}_1)^2 - (\nu'/\nu_{\mathrm{line}}-\tilde{g}_0)^2}} \non \, \\
    &\times \frac{e^{- 2 k_\mathrm{ps} \gamma_0}}{1-e^{-2\gamma_0}} \sum_{-m_\mathrm{max}}^{m_\mathrm{max}} e^{- |m| \gamma_0} G_{\ell_t}\left(T+m\tau_0\right) \\
    & \qquad \times G^\circ_{\ell_\mathrm{co}}\left(\pm_{\beta'} \arccos \frac{\nu'/\nu_{\mathrm{line}}-\tilde{g}_0}{a \sin \theta_o \tilde{g}_1} \mp_{\beta} \arccos \frac{\nu/\nu_{\mathrm{line}}-\tilde{g}_0}{a \sin \theta_o \tilde{g}_1} +m\delta_0 - \Omega_s(\tr_0) \left( T+m \tau_0 \right) \right) \notag \, ,
\end{align}
This result may be interpreted as follows. In the 3D space spanned by $(\nu,\nu',T)$ we obtain a correlation enhancement along the curves defined by the intersection of the hypersurfaces  
\begin{align}
    &T=-m\tau_0  \\
    &\pm_{\beta'} \arccos \frac{\nu'/\nu_{\mathrm{line}}-\tilde{g}_0}{a \sin \theta_o \tilde{g}_1} \mp_{\beta} \arccos \frac{\nu/\nu_{\mathrm{line}}-\tilde{g}_0}{a \sin \theta_o \tilde{g}_1} =-m\delta_0 \, , \label{eq:correlation ridge nu nu'} 
\end{align}
i.e. 1D ``ridges'' along which correlation is enhanced. Each choice of $(m,\pm_\beta,\pm_{\beta'})$ contributes a single ridge. Note that if the spin and the (small) inclination are known, it could be useful to integrate, in addition, over the coordinate $\pm_{\beta'} \arccos \frac{\nu'/\nu_{\mathrm{line}}-\tilde{g}_0}{a \sin \theta_o \tilde{g}_1} \pm_{\beta} \arccos \frac{\nu/\nu_{\mathrm{line}}-\tilde{g}_0}{a \sin \theta_o \tilde{g}_1}$ in the $\nu,\nu'$ plane (for some choice of $\pm_{\beta},\pm_{\beta'}$), keeping the LHS of Eqs.~\eqref{eq:correlation ridge nu nu'} fixed, in order to accumulate the ridge to a single correlation peak in the plane $\left(T,\pm_{\beta'} \arccos \frac{\nu'/\nu_{\mathrm{line}}-\tilde{g}_0}{a \sin \theta_o \tilde{g}_1} \mp_{\beta} \arccos \frac{\nu/\nu_{\mathrm{line}}-\tilde{g}_0}{a \sin \theta_o \tilde{g}_1} \right)$.

%%%%%%%%%%%%%%%%%%%%%%%%%%%%%%%%%%%%%%%%%%%%%%%%%%%%%%%%%%%%%%%%%%%%%%%%%%

\section{Observational considerations}\label{sec:observation}

Here we perform a preliminary, order-of-magnitude estimate of the signal-to-noise (SNR) ratio we can hope to achieve with present and future instruments. We will focus on the observable \eqref{eq:C 1D def}, the unresolved-STAC function, and assume a significant line emission\footnote{Unresolved-STACs could be relevant for VLBI as well, c.f. \cite{Brinkerink2021}, albeit with a broad emission spectrum.}. SNR estimates for the more fine-grained image-STAC functions \eqref{eq:C 3D def}, \eqref{eq:C 2D def}, which are relevant for resolved sources, will be considered elsewhere.

X-ray data that could be relevant for \eqref{eq:C 1D def} is recorded in the form of a list of arrival times and energies, or frequencies, for each recorded photon. The typical temporal resolution of such a list is $\sim 10 \mathrm{\mu sec}$. Its spectral resolution, for present missions, is of the order of $\sim 100$\,eV. The list of photons may be translated into a time series $f_\nu(t)$ with which \eqref{eq:C 1D def} is constructed. This time series is discretized in both time and frequency into bins, the size of which cannot be smaller than the instrument's temporal and spectral resolutions, respectively.
The next-generation observatory NewAthena is expected to deliver an $\sim$ order of magnitude improvement in performance specifications compared to current missions, in particular in spectral resolution and collecting area.

In order to estimate the SNR for $\mathcal{C}_{\nu \nu'}(T)$, defined in Eq.~\eqref{eq:C 1D def}, we add a zero mean noise component $n_\nu(t)$ to the signals which we assume to be uncorrelated across time and frequency. We can then write
\begin{align} \label{eq:signal + noise decomposition}
    \mathcal{C}_{\nu \nu'}(T) = \underbrace{\langle  \Delta f_\nu(t) \Delta f_{\nu'}(t+T) \rangle}_\mathrm{signal} + \underbrace{\langle  \Delta f_\nu(t) n_{\nu'}(t+T) + n_\nu(t) \Delta f_{\nu'}(t+T) + n_\nu(t) n_{\nu'}(t+T) \rangle}_\mathrm{noise} \,,
\end{align}
where we keep in mind that the angle brackets denote a time average, taken over a period $t_\mathrm{obs}$.
Now, we use the definition
\begin{align} \label{eq:SNR def}
    \mathrm{SNR} := \frac{\sigma(\mathrm{signal})}{\sigma(\mathrm{noise})} \,,
\end{align}
where $\sigma$ denotes the standard deviation. Moreover, we assume that at the order-of-magnitude level $\sigma(\Delta f_\nu(t))$ is frequency-independent, that the values of $\nu$, $\nu'$, and $T$ are such that a correlation ridge is attained, and that $\ell_t$, $\ell_\mathrm{co}$ are short enough such that lensing correlation dominates over local source correlation. Using these assumptions and dividing the numerator and denominator of \eqref{eq:SNR def} by $\sigma(n)$ which for simplicity is assumed to be frequency independent, we estimate for the $m=1$ correlation ridge,
\begin{align} \label{eq:SNR(C)}
\mathrm{SNR}(\mathcal{C}) \sim \frac{e^{-\gamma_0}\left[\mathrm{SNR}(\Delta f)\right]^2}{\sqrt{2\left[\mathrm{SNR}(\Delta f)\right]^2+1}} N_t^{1/2}  \,,
\end{align}
where $\mathrm{SNR}(\Delta f)=\sigma(\Delta f)/\sigma(n)$, and $N_t$ is the number of independent statistical realizations (in time) at a given frequency. Note that the SNR of higher-order correlation peaks $|m|>1$ may be estimated by replacing $e^{-\gamma_0} \to e^{-|m|\gamma_0}$ in Eq.~\eqref{eq:SNR(C)}. The factor $N_t^{1/2}$ arises since $\sigma(\mathrm{noise}) \sim N_t^{-1/2}$ while $\sigma(\mathrm{signal}) \sim \mathcal{O}(1)$. Note also that when $\mathrm{SNR}(\Delta f) \ll 1$, $\mathrm{SNR}(\mathcal{C}) \propto \mathrm{SNR}^2(\Delta f)$, while when $\mathrm{SNR}(\Delta f) \gg 1$, $\mathrm{SNR}(\mathcal{C}) \propto \mathrm{SNR}(\Delta f)$.

$\mathrm{SNR}(\Delta f)$ may be assessed in terms of the average number of photons per spectro-temporal bin of spectral width $\Delta \nu_\mathrm{bin}$ and temporal duration $\Delta t_\mathrm{bin}$. The dimensions of the bins may be varied within a range determined as follows. From below, they are bounded by the spectral/temporal resolutions of the instrument, respectively. An upper bound derives from the requirement that several bins should fit within a characteristic spectral/temporal separation between the primary and secondary correlation peaks; otherwise, separated correlation peaks cannot be observed. With this in mind, denoting the total photon count rate of the fluctuating part of the emission in a range $(\nu_\mathrm{min},\nu_\mathrm{max})$ by $n_\mathrm{ph}$\footnote{Note that with the above definition, $n_\mathrm{ph}$ depends on $\Delta t_\mathrm{bin}$; e.g., enlarging $\Delta t_\mathrm{bin}$ effectively integrates out some high frequencies and therefore suppresses $n_\mathrm{ph}$. This must be kept in mind when estimating $n_\mathrm{ph}$. It is instructive to note that the fluctuation power spectrum of AGN and XRBs in the low state is often reasonably well fit by a power law with index -2 over a wide frequency range, c.f. Fig.~1 of \cite{Gonzalez-Martin2012}.}, we may estimate the average number of photons per bin as $\sim n_\mathrm{{ph}} \Delta \nu_\mathrm{bin} \Delta t_{\mathrm{bin}}/(\nu_\mathrm{max}-\nu_\mathrm{min})$. We also add a multiplicative factor $f_\mathrm{source}$ in order to effectively account for signal suppression, relative to the overall flux, due to a specific source's structure. This includes scenarios where the line emission emanates far from the photon shell, source absorption cannot be neglected\footnote{In order to find correlations, a mostly optically thin source is required. On the other hand, in order to generate fluorescent iron emission, regions which are sufficiently optically thick in order to be efficiently irradiated are needed. Such a situation could be possible when the flow is highly inhomogeneous (clumpy), in which case one has pockets of neutral iron emitting regions embedded in a highly tenuous (Compton thin) gas component. This could be similar to the models qualitatively discussed by \cite{Dexter2011}.}, continuum emission swamps the line signal, and/or flow turbulence significantly smears the emission line in the GECO rest frame. It seems difficult to estimate $f_\mathrm{source}$ universally; its magnitude likely differs significantly between sources. Using these definitions we may estimate
\begin{align} \label{eq:SNR(delta f)}
    \mathrm{SNR}(\Delta f) \sim f_\mathrm{source} \left(\frac{n_\mathrm{{ph}} \Delta \nu_\mathrm{bin} \Delta t_{\mathrm{bin}}}{\nu_\mathrm{max}-\nu_\mathrm{min}}\right)^{1/2} \, .
\end{align}
The number of independent statistical realizations of the specific flux fluctuation at a fixed frequency bin can be estimated as $N_t \sim t_\mathrm{obs}/\mathrm{max\{\Delta t_\mathrm{bin},\ell_t\}}$.

These tools in hand, we will turn to estimate \eqref{eq:SNR(C)} for two different types of sources: XRBs and AGN. As an initial step, we use two particular examples to extract source parameters; a much wider survey is required for a better estimate of the SNR in general. We will begin with XRBs, taking as an example the bright source MAXI J1535-571 as observed by AstroSat/LAXPC. This BH's mass was estimated to be $(6.4 \pm 1.3) M_\odot$ in Ref.~\cite{Sreehari2019}, and therefore its typical time delay is $ \sim 15 \frac{GM}{c^3} \sim 2 \mathrm{ms}$. In this light, we should only consider $\Delta t_\mathrm{bin} \lesssim 1\mathrm{ms}$. The data show that in the $3-4\mathrm{keV}$ range, the photon count rate in the fluctuating part, quantified by the standard deviation, amounts to $n_\mathrm{ph}\sim 10^3 \mathrm{s^{-1}}$ at temporal bin size $\Delta t_\mathrm{bin} \sim 1\mathrm{ms}$. 
Thus, for this observation we obtain
\begin{align}
    \mathrm{SNR}(\Delta f) \sim \frac{1}{20} \left( \frac{f_\mathrm{source}}{0.1} \right)    \left(\frac{ \Delta \nu_\mathrm{bin}}{250\mathrm{eV}} \frac{ \Delta t_{\mathrm{bin}}}{1\mathrm{ms}}\right)^{1/2} \, .
\end{align}
Plugging this into \eqref{eq:SNR(C)}, assuming $\Delta t_\mathrm{bin} \gtrsim \ell_t$, a low photon count rate $\mathrm{SNR}(\Delta f) \ll 1$, and employing the corresponding approximation, we obtain
\begin{align} \label{eq:SNR(C) - XRB}
    \mathrm{SNR}(\mathcal{C}) \sim 3 \times 10^{-3} \left(\frac{e^{-\gamma_0}}{e^{-\pi}}\right) \left( \frac{f_\mathrm{obs}}{0.1} \right)^2 \left( \frac{\Delta \nu_\mathrm{bin}}{250\mathrm{eV}} \right) \left( \frac{\Delta t_\mathrm{bin}}{1\mathrm{ms}} \right)^{1/2} \left( \frac{t_\mathrm{obs}}{1 \mathrm{s}} \right)^{1/2} \, .
\end{align}
In order to obtain a rough estimate of the required observation time, we may substitute unity in all the factors in parenthesis in Eq.~\eqref{eq:SNR(C) - XRB} except the rightmost one where $t_\mathrm{obs}$ appears explicitly. Equating $\mathrm{SNR}(\mathcal{C})\sim 1$ and solving for the observation time implies that the signal can begin to rise above the noise when $t_\mathrm{obs} \gtrsim 1.3$ days. It is important to reiterate that significant assumptions have been made in order to obtain this estimate, in particular $f_\mathrm{obs} \sim 0.1$. Assuming, for example, $f_\mathrm{obs} \sim 0.01$, would have yielded a required observation time of $t_\mathrm{obs}\gtrsim35$~years.

Turning to AGN, we take as an example the nearby source I Zwicky 1, which was argued in Ref.~\cite{Wilkins2021} to show evidence of variable coronal emission reverberating off different parts of its accretion disk and displaying echoes. This BH's mass was estimated to be $(3.1 \pm 0.5) \times 10^7 M_\odot$ \cite{Wilkins2021}, and therefore its typical time delay is $\sim 30$ minutes. Taking its XMM-Newton observations as a test case, the data show a standard deviation of $\sim 0.07 \mathrm{s}^{-1}$ in the count rate fluctuations, in the $3-4$keV band. For this observation, we therefore obtain
\begin{align}
    \mathrm{SNR}(\Delta f) \sim 0.3 \left( \frac{f_\mathrm{source}}{0.1} \right)    \left(\frac{ \Delta \nu_\mathrm{bin}}{250\mathrm{eV}} \frac{ \Delta t_{\mathrm{bin}}}{500\mathrm{s}}\right)^{1/2} \, .
\end{align}
Plugging this into \eqref{eq:SNR(C)}, again assuming for simplicity $\Delta t_\mathrm{bin} \gtrsim \ell_t$ and a low photon count rate $\mathrm{SNR}(\Delta f) \ll 1$, gives
\begin{align}
    \label{eq:SNR(C) - AGN}
    \mathrm{SNR}(\mathcal{C}) \sim 2\times 10^{-4} \left(\frac{e^{-\gamma_0}}{e^{-\pi}}\right) \left( \frac{f_\mathrm{obs}}{0.1} \right)^2 \left( \frac{\Delta \nu_\mathrm{bin}}{250\mathrm{eV}} \right) \left( \frac{\Delta t_\mathrm{bin}}{500\mathrm{s}} \right)^{1/2} \left( \frac{t_\mathrm{obs}}{1 \mathrm{s}} \right)^{1/2} \, .
\end{align}
Estimating the required observation time for the signal to rise above the noise in a similar manner to that explained below Eq.~\eqref{eq:SNR(C) - XRB}, we obtain $t_\mathrm{obs} \gtrsim 0.8$ years.

To conclude this section, despite the existence of significant unknowns, we may take away several valuable lessons from the above analysis. First, it seems that XRBs seem to be somewhat preferred over AGN for the observation of unresolved STACs $\mathcal{C}_{\nu \nu'}(T)$. Second, that the SNR seems to improve significantly when the signal SNR obeys $\mathrm{SNR}(\Delta f) \gtrsim 1$, with an important role played by the photon count rate in that regard. Finally, that estimates of $f_\mathrm{source}$ could be valuable for a better assessment of the observability of unresolved STACs. 

We reemphasize that we have only considered in this section, as case studies, two particular observations of specific sources. There are many existing datasets corresponding to a variety of sources, observed by different missions, which could be considered. It is also important to reiterate the transformative capabilities of NewAthena in the present context. When it comes online, its collecting area is expected to increase that of e.g. XMM/PN by a factor of $\sim10$, together with other order-of-magnitude improvements. The above analysis implies that such improvements could have a significant effect on the SNR. 

\section*{Acknowledgements}

We thank A. Lupsasca for useful comments. We acknowledge financial support by the Data Science Research Center (DSRC) at the University of Haifa. S. Harikesh is supported by a Bloom postdoctoral fellowship. Research by D. Chelouche is partially supported by grants from the German Science Foundation (DFG HA3555-14/1, CH71-34-3) and the Israeli Science Foundation (ISF 2398/19). 

\appendix

\section{Explicit analytical expressions}
\label{app:explicit analytical expressions}

\subsection{Photon shell critical parameters}
\label{app:photon shell critical parameters}

\begin{subequations}\label{eq:explicit expressions for critical parameters}
The critical parameters $\{ \gamma,\delta,\tau \}$ are given by
\begin{gather}
    \gamma(\tilde{r}) = \frac{4\tilde{r}}{a\sqrt{-\tilde{u}_-}}\sqrt{1-\frac{M \tilde{\Delta}}{\tilde{r}(\tilde{r}-M)^2}} \, K\left( \frac{\tilde{u}_+}{\tilde{u}_-} \right) \, , \\
    \delta(\tilde{r}) = \frac{2}{\sqrt{-\tilde{u}_-}} \left[ \frac{\tilde{r}+M}{\tilde{r}-M} \, K\left( \frac{\tilde{u}_+}{\tilde{u}_-} \right) + \frac{\tilde{\lambda}}{a} \, \Pi\left( \tilde{u}_+,  \frac{\tilde{u}_+}{\tilde{u}_-} \right) \right] + 2\pi \Theta(\tilde{r}-\tilde{r}_0) \, , \\
    \tau(\tilde{r}) = \frac{2}{a \sqrt{-\tilde{u}_-}} \left[ \tilde{r}^2 \left( \frac{\tilde{r}+3M}{\tilde{r}-M} \right) \, K\left( \frac{\tilde{u}_+}{\tilde{u}_-} \right) - a^2 \tilde{u}_- \left( E\left( \frac{\tilde{u}_+}{\tilde{u}_-} \right) - K\left( \frac{\tilde{u}_+}{\tilde{u}_-} \right) \right) \right] \, ,
\end{gather}
\end{subequations}
where $K$, $E$, and $\Pi$ are, respectively, the complete elliptic functions of the first, second, and third kind. $\Theta$ represents the Heaviside function, and $\tilde{u}_\pm$ are defined in equation \eqref{eq:u pm definitions},
and are evaluated here for $\lambda = \tilde{\lambda}$, $\eta = \tilde{\eta}$.

\subsection{Redshift of plunging emitters}
\label{app:Redshift}

The redshift factor of geodesic emitters which plunge from the ISCO was derived in Ref.~\cite{Cunningham1975}. 
\begin{align}
g_{\mathrm{plunge}} = \frac{1}{u^t-u^\phi \lambda \mp u^r  \sqrt{\mathcal{R}(r_s)}/\Delta(r_s)} \,,
\label{eq:g for equatorial plunging geodesic emitters}
\end{align}
where the  $\mp=-\mathrm{sign}(p_s^r)$ corresponds to the radial direction of emission at the source, and
\begin{align}
&u^r = -\sqrt{\frac{2M}{3r_\mathrm{isco}}} \left(\frac{r_\mathrm{isco}}{r_s}-1\right)^{3/2} \,, ~~ u^\phi = \frac{\Gamma_\mathrm{isco}}{r_s^2} (\lambda_{\mathrm{isco}}+ a H)\,, ~~ u^t = \Gamma_\mathrm{isco} \left[ 1+\frac{2M}{r_s}(1+H) \right] \, , \non \\ 
&H = \frac{2M r_s-a\lambda_\mathrm{isco}}{\Delta(r_s)} \, ~~ \lambda_\mathrm{isco} = \frac{\sqrt{M}\left( r_\mathrm{isco}^2-2a\sqrt{M r_{\mathrm{isco}}}+a^2 \right)}{r_{\mathrm{isco}}^{3/2}-2M\sqrt{r_{\mathrm{isco}}}+a\sqrt{M}} \,, \Gamma_\mathrm{isco} = \sqrt{1-\frac{2M}{3r_\mathrm{isco}}} \,.
\end{align}
The ISCO is special in that (in the probe limit) three qualitatively different geodesics with the same value of orbital parameters exist: a marginally stable circular geodesic at $r=r_\mathrm{isco}$, a plunge from the ISCO, and its time-reversed ``climb'' to the ISCO.

\bibliography{STcorrelations.bib}
\bibliographystyle{utphys}

\end{document}